\documentclass[11pt]{article}
\usepackage[utf8]{inputenc}

\usepackage{amsmath,amsfonts,amssymb,amsthm}
\usepackage[margin=1in]{geometry}
\usepackage{graphicx,xcolor}
\usepackage{float}
\usepackage{physics}
\usepackage[colorlinks]{hyperref}
\usepackage[capitalize]{cleveref}

\usepackage{changes}

\usepackage{mathrsfs}
\usepackage{mathtools}
\mathtoolsset{centercolon}
\usepackage{bm}
\usepackage{multirow}
\usepackage{bbm}
\usepackage{subcaption}

\usepackage{tikz} 
\usetikzlibrary{arrows, shapes.gates.logic.US, calc}
\usetikzlibrary{shapes}
\usetikzlibrary{plotmarks}
\usetikzlibrary{quantikz}
\usetikzlibrary{patterns}
\usepackage{tabularx}
\usepackage{subcaption}

\usepackage[final]{pdfpages}

\newcommand{\Or}{\mathcal{O}}

\DeclareMathOperator{\polylog}{\mathrm{polylog}}
\DeclareMathOperator{\poly}{\mathrm{poly}}

\renewcommand{\d}{\mathrm{d}}

\newtheorem{problem}{Problem}
\newtheorem{theorem}{Theorem}
\newtheorem{lemma}{Lemma}

\theoremstyle{definition}
\newtheorem{definition}{Definition}

\allowdisplaybreaks

\renewcommand{\sec}[1]{\hyperref[sec:#1]{Section~\ref*{sec:#1}}}
\newcommand{\app}[1]{\hyperref[app:#1]{Appendix~\ref*{app:#1}}}
\newcommand{\thm}[1]{\hyperref[thm:#1]{Theorem~\ref*{thm:#1}}}
\newcommand{\lem}[1]{\hyperref[lem:#1]{Lemma~\ref*{lem:#1}}}
\newcommand{\cor}[1]{\hyperref[cor:#1]{Corollary~\ref*{cor:#1}}}
\newcommand{\prb}[1]{\hyperref[prb:#1]{Problem~\ref*{prb:#1}}}
\newcommand{\fgr}[1]{\hyperref[fgr:#1]{Figure~\ref*{fgr:#1}}}
\newcommand{\tab}[1]{\hyperref[tab:#1]{Table~\ref*{tab:#1}}}

\newcommand{\beq}{\begin{equation}}
\newcommand{\eeq}{\end{equation}}
\newcommand{\beqa}{\begin{eqnarray}}
\newcommand{\eeqa}{\end{eqnarray}}

\title{Toward end-to-end quantum simulation for protein dynamics}

\author{Zhenning Liu$^{1,2}$, Xiantao Li$^{3}$, Chunhao Wang$^{4}$, Jin-Peng Liu$^{5,6,\thanks{liujinpeng@tsinghua.edu.cn}}$ \\ 
\footnotesize $^{1}$ Joint Center for Quantum Information and Computer Science, University of Maryland\\
\footnotesize $^{2}$ Department of Computer Science, University of Maryland\\
\footnotesize $^{3}$ Department of Mathematics, Pennsylvania State University\\
\footnotesize $^{4}$ Department of Computer Science and Engineering, Pennsylvania State University\\
\footnotesize $^{5}$ Yau Mathematical Sciences Center, Tsinghua University\\
\footnotesize $^{6}$ Beijing Institute of Mathematical Sciences and Applications\\
}

\date{}

\begin{document}

\maketitle

\begin{abstract}
Modeling and simulating the protein folding process overall remains a grand challenge in computational biology. We systematically investigate end-to-end quantum algorithms for simulating various protein dynamics with effects, such as mechanical forces or stochastic noises. 
A major focus is the read-in of system settings for simulation, for which we discuss (i) efficient quantum algorithms to prepare initial states—whether for ensemble or single-state simulations, in particular, the first efficient procedure for preparing Gaussian pseudo-random amplitude states, and (ii) the first efficient loading of the connectivity matrices of the protein structure. 
For the read-out stage, our algorithms estimate a range of classical observables, including energy, low-frequency vibrational modes, density of states, displacement correlations, and optimal control parameters. 
Between these stages, we simulate the dynamic evolution of the protein system, by using normal mode models—such as Gaussian network models (GNM) and all-atom normal mode models. 
In addition, we conduct classical numerical experiments focused on accurately estimating the density of states and applying optimal control to facilitate conformational changes. These experiments serve to validate our claims regarding potential quantum speedups. Overall, our study demonstrates that quantum simulation of protein dynamics represents a robust, end-to-end application for both early-stage and
fully fault-tolerant quantum computing.
\end{abstract}

\section*{Introduction}\label{sec:introduction}

The precise functions of biological systems frequently hinge on the ability of proteins to undergo conformational transitions. These structural changes encompass a broad range of movements --- from minor adjustments of individual side-chains and loops to extensive shifts involving large domains \cite{schlick2010molecular}. To thoroughly understand and predict these dynamic processes, it is crucial to simulate the motions of protein atoms over biologically relevant timescales, enabling the identification of key atomic properties that dictate protein function \cite{klepeis2009long,lindorff2011fast},  e.g., in target-based drug designing, elucidating protein-related disease mechanisms, and innovating novel enzymes \cite{maria2018role}.
However, current molecular dynamics simulations still face significant challenges in this domain. While they are effective for exploring small-scale transitions, the computational demands of these methods become prohibitive when studying more extensive molecular motions. This limitation stems from the inadequate sampling of conformational space within the typical simulation durations of just tens of nanoseconds, which are insufficient to capture large-scale dynamics. Moreover, the complexity of protein structures, characterized by numerous local minima due to their large number of atoms, further complicates the simulation process.
Thus, modeling and simulating the protein folding process overall remains a grand challenge in computational biology. While recent advancements in AI-based methods such as AlphaFold3~\cite{jumper2021highly} have been significant, their precision in determining the structure of proteins with low sequence similarity is still limited.

%Computational techniques have been extensively developed to simulate the molecular dynamics (MD) to enhance computational efficiency~\cite{frenkel2023understanding,schlick2010molecular}. However, the considerable disparity between the time scales of molecular movements ($10^{-15}$s) and those of conformational changes (at least $10^{-3}$s), coupled with the vast number of atoms involved, necessitates massive computing power to simulate protein dynamics~\cite{klepeis2009long} directly.
%\xl{Upon re-reading, I realize that the content of this paragraph is a repetition of what was already conveyed in the preceding one.   }
%\jpl{agreed to comment out this paragraph}

Recent advances in quantum computing offer a new paradigm for large-scale computing.  On the study of protein structure problems, prior works mainly focused on mapping the coarse-grained protein folding problems onto unconstrained or constrained optimization problems, for which they explored near-term algorithms such as quantum annealing algorithms, Varitional Quantum Eigensolver (VQE), and Quantum Approximated Optimization Algorithm (QAOA) to obtain heuristic solutions~\cite{babbush2014construction,fingerhuth2018quantum,robert2021resource,irback2022folding,linn2023resource,bopardikar2023approach,doga2023perspective}. While these algorithms are feasible for current Noisy Intermediate-Scale Quantum (NISQ) devices, a provable quantum speedup has yet to be established. 

Meanwhile, quantum algorithms with provable advantage have been widely developed to simulate quantum dynamics, as well as to solve general linear and certain nonlinear differential equations on both early fault-tolerant and fully fault-tolerant quantum devices.  More importantly, the goals of these algorithms are more aligned with those of protein dynamics simulations. Notably, these approaches provide the quantum states that encode the solution of high-dimensional dynamical systems, only requiring complexity poly-logarithmic in dimension, which can offer potential exponential speedups over classical simulation algorithms. A large variety of quantum algorithms have been developed for Hamiltonian simulations~\cite{Llo96,BAC07,BCC13,BCC15,LC16,LC17}, solving systems of linear equations~\cite{HHL08,Amb12,CKS15,AL19,LT19} and differential equations~\cite{Ber14,BCOW17,CL19,CLO20,LKK20,ALL23,ALWZ22}, as well as simulating open quantum systems~\cite{KBG11,CL17,CW17,LW23,LW23a,CKBG23,DLL24,PSW23}.

%\ZL{I've changed everything here to capital $N$ to make it consistent with the input model chapter. Please let me know if this may cause any difficulty.}\jpl{Looks good to me}
The dynamics of proteins can be accurately modeled by Newton's equations \cite{frenkel2023understanding}. A simplified, yet still extremely useful model \cite{bahar2010normal}, is  the normal mode dynamics under the mechanical force and/or stochastic noise
\begin{equation}\label{eq:md}
    M\frac{\d^2\bm u}{\d t^2} = - \gamma \frac{\d\bm u}{\d t} - K\bm u + \bm f(t) + \bm \xi(t).
\end{equation}
For a model of $N$ components, $\bm{u}\in \mathbb{R}^N$ contains the displacements;
%\ZL{Is this the same as the number of $C^\alpha$ atoms? Or should we say ``displacement of components"?}
%\jpl{Can we just say it is the displacement? The displacement of the atoms or molecules can also work for me}
%\ZL{I think we need to explicitly say what is $N$}
%\ZL{How about this?}\jpl{Good to me}
$M\in \mathbb{R}^{N\times N}$ is a diagonal matrix that contains the mass of the components, and the stiffness matrix $K\in \mathbb{R}^{N\times N}$ represents the interactions in the protein. $\gamma$ is a friction coefficient. $\bm f(t)$ can be regarded as a mechanical load, and $\bm \xi(t)$  is a stochastic force emanating from the environment, e.g., solvent. The normal mode analysis (NMA) has demonstrated its ability in efficiently predicting vibration modes that correlate well with large-scale conformation changes of proteins \cite{bahar1997direct}, and has played a key role in many simulation-based studies as well as coarse-graining methods \cite{li2002coarse,tama2000building,durand1994new}.

We will demonstrate that the dynamics in \cref{eq:md} can be reformulated as quantum dynamics that are efficiently simulatable on a quantum circuit. However, to achieve a full overall quantum advantage, significant challenges remain, including the development of protocols to encode proteins' initial state and connectivity, as well as measurement tools to extract quantities of interest. Recent work \cite{BBK23} has explored quantum simulation of high-dimensional coupled oscillator systems, highlighting potential applications to classical dynamical systems. Nevertheless, for many practical models of protein dynamics -- which may involve non-Hermitian systems, stochastic forces, and various target output -- new quantum algorithms are needed. {In addition, Ref.~\cite{BBK23} assumed efficient initial state preparation and access to the coupled oscillator system setting, which may not be available to practical protein dynamics simulations.}

\textbf{Contribution:}  We systematically investigate end-to-end quantum algorithms for simulating protein molecular dynamics that incorporate non-Hermitian structures as well as external forces and noise. A major focus is the read-in of system settings for simulation, for which we discuss efficient loading of the connectivity matrices of the protein structure for the first time and design quantum algorithms to prepare initial states—whether for ensemble or single-state simulations. In particular, we design the first efficient procedure for preparing Gaussian pseudo-random amplitude states.  For the read-out stage, our algorithms estimate a range of classical observables, including energy, low-frequency vibrational modes, density of states, displacement correlations, and optimal control parameters.
Between these stages, we simulate the dynamic evolution of the protein system, by using normal mode models—such as Gaussian network models (GNM) and all-atom normal mode models, by developing efficient quantum simulation algorithms that generate quantum encodings of the protein configuration at a given time \( T \), quantum history states for the trajectories, or covariant matrices for inhomogeneous or stochastic harmonic oscillator models. In addition, we conduct classical numerical experiments focused on accurately estimating the density of states and applying optimal control to facilitate conformational changes. These experiments serve to validate our claims regarding potential quantum speedups. Overall, our study demonstrates that quantum simulation of protein dynamics represents a robust, end-to-end application for both early-stage and fully fault-tolerant quantum computing.

\section*{Results}
To precisely compare the computational efficiency with classical simulation algorithms, we determined the number of qubits and the circuit depth (time complexity) required to prepare and simulate protein dynamics. For clarity, these results are presented in the form of theorems.

\subsection*{I.Theorems}
%\ZL{Do we need a better way to refer to the supplemental materials? Can we give all sections in the supplement a letter S, such that we can simply say Section S4?}\jpl{Good idea. I have renamed the sections in the supplement materials like S4}
%\jpl{Zhenning: you can add around 800-1000 words for initial state and matrix loading sections}

%\ZL{working here... still thinking about which part to include. sorry for the delay}

\textbf{(i) Initial state preparation in Gaussian network model:}

To simulate the dynamics of the protein dynamical system with $N$ amino acid residues, our quantum strategy uses $n=\Or(\log N)$ qubits and employ some version of \emph{amplitude encoding}, i.e., we
%Regardless of which model we use,
%To achieve quantum computational advantage, we
%employs some sort of amplitude encoding of the dynamical state,
%the initial quantum state should be the amplitude encoding of the initial dynamical state,
encode the relative displacement and velocity of every $C^\alpha$ atom in the system as the amplitude corresponding to the atom's index. In this paper, we adapt the algorithm suggested in Ref.~\cite{BBK23} to perform efficient simulation of coupled oscillator dynamics. If $\bm{u}(t)$ and $\dot{\bm{u}}(t)$ are the relative displacements and velocities of the atoms at time $t$, then Ref.~\cite{BBK23} uses
\begin{equation}\label{psi-encode}
    \ket{\psi(t))} \coloneqq \frac{1}{\sqrt{2E}} [\dot{\bm y}(t), \mathrm{i} \sqrt{A}\bm y(t)]^T
\end{equation}
to encode them, where $\bm y(t) = \sqrt{M} \bm u(t)$ and $A \coloneqq (\sqrt{M})^{-1}K(\sqrt{M})^{-1}$. At $t=0$, the initial state $\ket{\psi(0)}$ represents the initial configuration of the dynamical system, $\bm{u}(0)$ and $\dot{\bm{u}}(0)$.
%high-dimensional coupled oscillator system of the corresponding model of protein.
%Ref.~\cite{BBK23} requires the initial state $\ket{\psi(0)}$ to be
%\begin{equation}
%    \ket{\psi(0))} \coloneqq \frac{1}{\sqrt{2E}} [\dot{\bm y}(0), i \sqrt{A}\bm y(0)]^T,
%\end{equation}
%where $\bm y(0) = \sqrt{M} \bm u(0)$ and $A \coloneqq (\sqrt{M})^{-1}K(\sqrt{M})^{-1}$.
For $t\geq 0$, we will demonstrate that $\ket{\psi(t)}$ satisfies an equation analogous to the Schr\"odinger equation. Consequently, once $\ket{\psi(0)}$ is prepared, we can employ a quantum algorithm to generate $\ket{\psi(t)}$ for any $t>0$, extending the result in Ref.~\cite[Lemma 9]{BBK23}.
%Note that $\ket{\psi(0)}$ (and $\ket{\psi(t)}$) . 

%Observe that initial displacements and initial velocities are represented by amplitudes of non-overlapping sets of values in the register. We can show that if we are able to prepare the states proportional to the initial displacements and initial velocities respectively and are given oracle access to both the $M$ and $K$ matrices, then the initial state can be prepared efficiently.

%The initial state $|\psi(0)\rangle$ needs to be proportional to the initial displacements and velocities of all $N$ oscillators in the system. 

The \emph{joint state preparation} lemma (see S5.1 of Supplemental Materials for details) from Ref.~\cite{BBK23} states that if one can efficiently implement unitary $U_{I}$ such that $U_{I}\ket{0}\ket{0} =\ket{0}\ket{\dot{\bm u}(0)}, U_{I}\ket{1}\ket{0} = \ket{1}\ket{\bm u(0)}$ where $\ket{\bm u(0)} \propto \sum_{j=1}^N u_j(0)\ket{j}$ and $\ket{\dot{\bm u}(0)}  \propto \sum_{j=1}^N \dot{u}_j(0)\ket{j}$, then $\ket{\psi(0)}$ can be prepared efficiently. Since the implementation of $U_I$ was not elaborated in Ref.~\cite{BBK23}, we investigate the construction of $U_{I}$ here.
%, which may be different for different protein dynamics models.
In particular, we focus the Gaussian network mode, where every initial displacement $u_j(0)$ and velocity $\dot{ u}_j(0)$ follows a Gaussian distribution $\mathcal{N}(0,\mu)$ independently. %Therefore, 
%all coefficients of the displacement state $\ket{\bm u(0)}$ and velocity state $\ket{\dot{\bm u}(0)}$ should be random variables with zero mean, and 
%$\bm u_j(0)$ and $\dot{\bm u}_j(0)$ independently follows a real Gaussian distribution $\mathcal{N}(0,\mu)$ 
We call the corresponding quantum state $\ket{\dot{\bm u}(0)} $ and $\ket{\bm u(0)}$ \emph{Gaussian random-amplitude state}. $U_I$ should simply generate Gaussian random-amplitude states.

We identify two classes of initial state preparation problems for two types of computational tasks related to protein dynamics simulation. The first kind of tasks, to be referred to as the \emph{ensemble simulation}, is computing the averaged behavior of the system over the distribution of all possible input states. In this case, the initial state should be \emph{sampled} from the Gaussian distributions, as described in the following problem.
\begin{problem}[Initial state preparation for ensemble simulation in GNM]
\label{prb:ensemble}
    Use $\poly(n)$ gates and qubits to prepare a mixed state $\rho_\mathrm{G}$ defined as the convex combination of all $n$-qubit Gaussian random amplitude states weighted by their probability densities.
\end{problem}
This problem turns out to have a straightforward solution as follows.
\begin{theorem}
\label{thm:ensemble}
    $\rho_\mathrm{G}$ can be prepared using a depth-2 circuit on $2n$ qubits.
\end{theorem}
Please see S5.2 of Supplemental Materials for the proof. In fact, $\rho_\mathrm{G}$ is just the maximally mixed state, which is easy to prepare: simply discard the second register of
$\ket{\psi_\mathrm{pure}} := \frac{1}{\sqrt{2^n}}\sum_{j=0}^{2^n-1}\ket{j}\ket{j}$, where $\ket{\psi_\mathrm{pure}}$ can be prepared by performing Hadamard gates on all qubits in register 1 and applying CNOT gates between register 1 and register 2. Interestingly, $\ket{\psi_\mathrm{pure}}$ itself is also useful as it can enable a quadratic speedup in estimating some observables through quantum mean estimation \cite{BHM02,cornelissen2022near,hamoudi2021quantum} (see S5.3 of Supplemental Materials for more details).

An in-depth study of the protein dynamical system under one specific random initial condition can also be useful for biology and chemistry. For instance, one may be interested in the evolution of a single initial configuration, which requires multiple instances of simulations under the same initial state, e.g., under different realizations of the noise $\bm \xi(t)$. One may also study how the properties and dynamics of the system change when a small perturbation is applied to the initial configuration, which requires the capability of fine-tuning the state preparation procedure to produce a slightly different initial state deterministically.
These features above may also be useful when studying a class of randomized simulation and their path dependence. We call this situation \emph{single-state simulation}, where the initial state is one specific reproducible Gaussian random (or pseudo-random) amplitude state sampled from its distribution.
%We define the initial state preparation problem for single-state simulation.

\begin{problem}[Initial state preparation for single-state simulation in GNM]
\label{prb:single}
    Prepare one Gaussian random amplitude state 
    %(representing $\ket{\bm{u}(0)}$ or $\ket{\bm{\dot{u}}(0)}$) 
    of $n$ qubits (pseudo-) randomly using $\poly(n)$ gates and qubits. We also need to reproduce this specific Gaussian random amplitude state arbitrarily many times.
\end{problem}

%To solve Problem~\ref{prb:single}, we need to generate specific $\ket{\psi_{\bm{\alpha}}}$ states where $\bm{\alpha}$ satisfies the Gaussian randomness condition and normalization condition.
Note that if the random state is generated probabilistically by measuring a larger quantum state, then it might be difficult to reproduce the same state or a slightly modified state due to the no-cloning theorem and the large overhead of post-selection. Therefore, we prefer to produce the randomness using deterministic pseudo-random generators, thus the state becomes pseudo-random rather than true random \footnote{Note that true randomness is not necessary for our purpose because we only aim to achieve the same outcomes of protein dynamics simulation as classical approaches, and scientists only have access to pseudo-randomness with classical computers. %ZL{Maybe add a reference here...}
}. The most conceptually simple method is to sample a Gaussian random array $\bm{\alpha}$ classically and use a quantum circuit to prepare $\ket{\bm \alpha} \propto \sum_{j=1}^N \alpha_j \ket{j}$ accordingly. However, since there are $\Or(2^n)$ bits in $\bm{\alpha}$, the state preparation process is inefficient. To address this problem, we devise a new algorithm for preparing Gaussian pseudo-random amplitude states based on the Gaussian random amplitude loading approach introduced in Section 4.1 of~\cite{BDP23}.

\begin{theorem}[Gaussian pseudo-random amplitude state preparation]
\label{thm:pseudorandom}
    A single Gaussian pseudo-random amplitude state of $n$ qubits can be generated deterministically using $\poly(n)$ quantum gates and $\poly(n)$ classical bits which are used as the source of randomness.
\end{theorem}

%\ZL{maybe point to an appendix for detailed proof}
We briefly describe our techniques as follows and leave details to S5 of Supplemental Materials. Our approach uses binary data loaders extensively, the same as in Ref.~\cite{BDP23}. A binary data loader $L_h$ parametrized by angles $\bm{\theta} = (\theta_0,\theta_1,\dots,\theta_{2^n-1})$ acting on the $l$th qubit implements the following operation for $i\in\{0,1\}^{l-1}$:
\begin{equation}
    L_l(\bm{\theta}):\ket{i}\ket{0} \rightarrow \cos(\theta_i)\ket{i}\ket{0} + \sin(\theta_i)\ket{i}\ket{1}
\end{equation}
%where $i$ is an integer represented by a binary string of $l-1$ bits, and $\theta_i \in [0,\pi/2]$.
%In fact, any $n$-qubit state $\ket{\phi}=\sum_j \beta_j \ket{j}$ with $\beta_j \in \mathbb{R}$ can be prepared using $n$ binary data loaders: if $\theta_i$ is computed by
%\begin{equation}
%    \cos(\theta_i) =\frac {\sum_{j\in \{0,1\}^{n-l}} |\beta_{i0j}|^2}{\sum_{k\in \{0,1\}^{n-l+1}} |\beta_{ik}|^2},
%\end{equation}
%then one can prepare $\ket{\phi}$ by applying each $L_l$ with $l\in\{1,2,\dots,n\}$ sequentially. Since $\bm{\theta}$ has $\Or(2^n)$ elements, it is expensive to compute all $\theta_i$ values classically and communicate them to the quantum circuit.
%We instead choose to generate $\bm{\theta}$ directly by the quantum circuit itself.

For Gaussian pseudo-random amplitude states, the following lemma from Ref.~\cite{BDP23} gives a hint for how to generate these $\theta_i$ angles.
\begin{lemma}[Lemma 4.8 of~\cite{BDP23}]
    The $n$-qubit quantum state $\ket{\psi_{\bm{\alpha}}}$ where $\alpha_i$ are independent $\mathcal{N}(0, 1)$ random variables can be prepared using the binary data loader construction with angles $\theta_i$ at the $l$th qubit distributed independently according to the density function 
    $$p_l(\theta_i) \propto \sin^{2^{n-l}-1}(\theta_i) \cos^{2^{n-l}-1}(\theta_i) \propto \sin^{2^{n-l}-1}(2 \theta_i).$$
\end{lemma}

We need to sample all $\theta_i$ angles according to the distribution above to produce one Gaussian random amplitude state. The approach in Ref.~\cite{BDP23} is preparing another state
\begin{equation}
    \ket{\psi_{\bm{\theta}}} \propto \sum_{\bm{\theta}} \sqrt{\prod_{l=1}^n \prod_{i\in\{0,1\}^{l-1}} p_l(\theta_i)} \ket{\bm{\theta}}
\end{equation}
using the Grover-Rudolf algorithm \cite{GR02} in $\Or(n)$ depth. However, this method gives true random states, does not allow for reproducing the same random state, and is expensive because it takes $\Or(2^n)$ qubits to store the whole $\bm{\theta}$. 

We propose a new algorithm to address the issues above: for each $l$, sample $\theta_i$ \emph{pseudo-randomly} for all $i$ with $l-1$ bits \emph{coherently} before applying the $l$th binary data loader. To do so, we design a reversible classical circuit for generating a pseudo-random sample of $\theta_i$ 
%for any $i$ of $l-1$ bits, 
\emph{without} knowing the value of any $\theta_j$ with $j\neq i$.
The reversible circuit can be naturally quantized by replacing all classical gates with quantum counterparts, which can generate all $\theta_i$s for the same binary data loader in superposition.
%, and then translate the circuit into a quantum one by replacing classical gates with quantum gates.

As an important building block of our approach, we first introduce counter-based pseudo-random number generators (CBRNG).
%some non-recursive random number generators.
\begin{definition}[CBRNG]
An $n$-bit CBRNG is an efficiently computable (i.e., in $\poly(n)$ time) function which satisfies $\mathrm{CBRNG}(s,i) = r(s,i)$ for all $i\in\{0,1,\dots, 2^n - 1\}$, where $s\in\{0,1,\dots, 2^{\poly(n)}-1 \}$ is a random seed, and $r(s,i)$ is a pseudo-random number sampled from the uniform distribution of $\{0,1,\dots,\max(r)\}$ with $\max(r) \leq 2^{\poly(n)}$.
\end{definition}
An example of CBRNG is the \emph{counter mode of block cipher} method, whose basic idea is to encode the index $i$ by permuting its bits corresponding to the seed.  More CBRNG proposals can be found in Refs.~\cite{salmon2011parallel,widynski2020squares}. Since CBRNGs are classical deterministic procedures, one can always produce the same $r(s,i)$ values if a fixed seed $s$ is chosen.

%We present both algorithms of generating $\bm{\theta}$ as follows.

%\paragraph{Algorithm 1 (naive).} CBRNG enables a straightforward (but not necessarily efficient) sampling method for $\theta_i$s based on inverting the cumulative distribution function, which is described as follows.

Without loss of generality, let us consider the data loader acting on the $l$th qubit, which has $2^{l-1}$ possible input values, i.e., the state at this step is $\ket{\psi_l} := \sum_{i\in\{0,1\}^{l-1}} \alpha_{l,i} \ket{i}$ with $\sum |\alpha_{l,i}|^2 = 1$. For every $i$ in this superposition, we need to generate its corresponding $\theta_i$ simultaneously.
First, we choose a random seed $s$ and compute $r(s,i)$ using the CBRNG for each $i$ in superposition, i.e.,
%More specifically, let the state prior to the $l$th binary data loader be $\sum_{i\in\{0,1\}^{l-1}} \alpha_{l,i} \ket{i} $, then the (quantized) CBRNG gives
\begin{equation}
   \sum_{i\in\{0,1\}^{l-1}} \alpha_{l,i} \ket{i} \ket{0} \mapsto \sum_{i\in\{0,1\}^{l-1}  }  \alpha_{l,i} \ket{i}\ket{r(s,i)}.
\end{equation}
Then, $\theta_i$ can be computed from $r(s,i)$ if we can \emph{invert} the cumulative distribution function (cdf) of $\theta_i$, i.e. computing the value $\theta_i$ such that
$\int_{0}^{\theta_i} p_l(x) dx = \frac{r(s,i)}{\max(r)}$. However, it is unclear if the above integration can be efficiently computed for $l\ll n$.
%thus inverting the cumulative distribution function might require numerical integration approaches and Newton's iterative method. We leave it as an open problem to find and quantize the best method for inverting the cumulative distribution function and analyzing its resource requirements.
To circumvent the hardness of integrating and inverting the cdf, we use the classical \emph{rejection sampling} method to generate $\theta_i$:
%Rejection sampling is a widely used approach in classical computation to sample from hard-to-sample-from distributions, which we outline as follows.
for a target distribution $p_l(x)$, we need to find a proposal distribution $q_l(x)$ which is easy to sample from
%(for instance, the cumulative distribution function of $q(x)$ may be efficiently invertible)
and a real number $Q_l$ such that $Q_l q_l(x) \geq p(x)$ for all $x$. Then, we generate $s_q$ from $q_l(x)$ and $r_q$ uniformly from $[0,Q_l q_l(s_q)]$. Next, if $r_q\leq p_l(s_q)$, then $s_q$ is accepted as a sample from $p_l(x)$; otherwise, $s_q$ is abandoned, and the we generate a new $(s_q,r_q)$ pair to repeat the above process. The success probability of obtaining one sample using this approach is $p_{\mathrm{suc},l} := \frac{\int p_l(x) dx}{\int Q_l q_l(x) dx}$, therefore, it is key to choose an appropriate $Q_l q_l(x)$ that it is sufficiently close to $p_l(x)$.

Indeed, we can show that such a proposal distribution exists. We notice that for $n-l \lesssim \log n$, sampling from $p_l(\theta_i)$ is not difficult, since $\int p_l(x) dx$ can be written analytically using $\poly(n)$ terms. Therefore, we focus on the case where $n-l \gg \log n$ and prove the following lemma (see S5.4 of Supplemental Materials for the proof).
\begin{lemma}[Proposal distribution]
    For $ l':= n- l \gg \log n$, there exists a $Q_l \in \mathbb{R}$ and a proposal distribution $q_l(x)$ for every $l<n$ such that rejection-sampling success probability satisfies $p_{\mathrm{suc},l} = \Or(1)$ and $q_l(x)$ can be efficiently sampled from by inverting its cumulative distribution function using pseudo-random numbers.
\end{lemma}

With the proposal distribution, one can show that rejection sampling can be implemented as a quantum procedure straightforwardly. We postpone the technical details of completing this quantized rejection sampling procedure to S5.5 of Supplemental Materials.

\subsubsection*{(ii) Protein structure matrix connectivity:}

%We describe how to implement the oracle to load the stiffness matrix $K$ onto the quantum system. $K$ encodes all $\Omega(N)$ harmonic oscillators in the protein. We point out that this is a particularly difficult problem in all quantum simulation of similar types of dynamical systems, since all information in $K$ are provided by scientific observations and cannot be compressed to $O(\log N)$ classical bits due to information theoretic limits. Therefore, our data loading approach is based on quantum read-only memory (QROM) which can be implemented in $O(\log N)$ depth but uses $O(N\log N)$ gates. We present how to implement the access oracle for $d$-sparse $K$ by loading protein atom or residue positions via QROM. Besides the brute-force sparse access, we also design a flexible binary tree data structure via QROM. This can be used to efficiently modify the protein structure, including changing the masses or positions of residues and adding or removing atoms or residues, by only modifying $O(d\polylog(N))$ entries of the QROM, rather than $O(d^2)$ in the brute-force case.

Previous work of quantum simulation of coupled oscillators \cite{BBK23} assumed that matrix $K$ can be accessed through a quantum \emph{sparse}-access oracle. %which can be implemented by $O(1)$ or $\polylog (N)$ effort.
A sparse-access oracle with query $(i,j)$ gives the value of the $j$th non-zero element in the $i$th row of the matrix. This is a standard assumption made in most existing quantum algorithms for differential equations literature, which mainly consider how many times the oracle is \emph{queried}, i.e., query complexity. In contrast, we aim at showing end-to-end quantum advantage in this paper, thus, we need to implement the data access oracle as a quantum circuit. Ideally, the circuit should consist of $\Or(\polylog N)$ gates and take $\Or(\polylog N)$ total effort to construct. However, as we prove in Theorem S1.10 in Supplemental Materials, such an efficient construction is only possible when the minimal number of bits to describe the $K$ matrix is $\polylog (N)$. Unfortunately, the $K$ matrix for protein molecular dynamics simulation does not satisfy the $\polylog(N)$-bit condition, as discussed in the following paragraph.

%Recall that $K_{i,j}$ is the spring constant between two $C^\alpha$ atoms, $i$ and $j$. 
Although the number of $(i,j)$ pairs is $\Or(N^2)$, a realistic molecule model typically has some locality such that the number of non-zero $K_{ij}$ is $\Or(N)$. Indeed, as presented in S2.2 of Supplemental Materials, the connection between any two $C^\alpha$ atoms in GNM or the anisotropic network model (ANM) is determined by the cutoff radius and the Euclidean distance between them. \footnote{Note that in the harmonic $C^\alpha$ Potential model (see S2.2.3 of Supplemental Materials), all pairs of $C^\alpha$ atoms have non-zero interactions, so the $K$ matrix is dense.} In both models, $K_{i,j} = f_K (|\vec{x}_i-\vec{x}_j|)$, where $f_K$ is an efficiently computable function, $\vec{x}_i$ and $\vec{x}_j$ are coordinates of the $i$th and the $j$th $C^\alpha$ atoms, respectively. Therefore, to load the $K$ matrix, it suffices to load the positions of $C^\alpha$ atoms and compute the $f_K$ function. Coordinates of $C^\alpha$ atoms partially determine the molecular structure, which is usually obtained by structural biology research \cite{yip2020atomic,jumper2021highly} instead of being generated by an efficient computer program. Therefore, the protein structure information has high information entropy and cannot be compressed with a ratio better than a constant value. In fact, even constant-ratio compression may be difficult, since typical molecular dynamics simulation tasks need high numerical accuracy, which requires lossless loading of atomic coordinates.
In conclusion, the matrix $K$ is incompressibly determined by the coordinates of $C^\alpha$ atoms, and since the number of atoms is $\Or(N)$, we need to load $\Or(N)$ bits.

%\ZL{Jin-Peng: please have a look at this paragraph and see if you are happy with the arguments to justify expoenential speedup.}\jpl{looks good to me}
Due to the fundamental limit presented above, we instead focus on $\Or(\polylog N)$-\emph{depth} implementation of the data access oracle.
%for potential exponential speedup in time complexity.
To achieve it, we employ a circuit-based quantum read-only memory (QROM), the quantum counterpart of classical read-only memory (ROM), using $\Omega(N)$ gates (see S1.3 of Supplemental Materials for a formal introduction, and S4 for our novel way of constructing low-depth QROM by quantization of a ROM).
We notice that our QROM and ROM share the same architecture and both require $\Omega(N)$ gates, but the cost of querying a ROM is generally considered as negligible in classical computer science research. Therefore,
%we believe that 
in the fault-tolerant era of quantum computing, it is reasonable to treat QROM in the same way as classical ROM, and the improvement in circuit depth should allow for a futuristic exponential speedup in time complexity with a practical application.
\begin{problem}[Low-depth matrix connectivity loading]
\label{prb:lowdepthmcl}
    Build a data access circuit (oracle) to load the matrix $K$ onto the quantum algorithm in $\Or(\polylog N)$ depth.
\end{problem}
Recall that the algorithm in Ref.~\cite{BBK23} requires the sparse-access oracle, which can be implemented if the following oracles are available:
\begin{equation}
\begin{aligned}
U_K \ket{i}\ket{j} =& \ket{i}\ket{j}\ket{K_{i,j}},\\
    U_s \ket{i}\ket{k}\ket{0} =& \ket{i}\ket{k}\ket{j_{i,k}},    
\end{aligned}
\end{equation}
where $j_{i,k}$ 
%\chw{the use of $j$ is conflict with the index $j$. Use another letter?}\jpl{Zhenning:how about $c_{i,k}$/$r_{i,k}$ which mean column/row index?}\ZL{but $i$ is already the row index, meaning that $r_{i,k}=i$ and $c_{i,k}=c_{r,k}$. I think the current version should be fine because $i,j$ also represents the atom index ($K_{i,j}$ is the spring between atom $i$ and atom $j$.)} 
is the column index of the $k$th non-zero element of the $i$th row of $K$. Note that $U_K$ can be implemented using the following 2-step strategy:
\begin{equation}
\ket{i}\ket{j}\ket*{\vec{0}}\ket*{\vec{0}}\ket{0} \xrightarrow{U_x\otimes U_x} \ket{i}\ket{j}\ket{\vec{x}_i}\ket{\vec{x}_j}\ket{0} \xrightarrow{U_f} \ket{i}\ket{j}\ket{\vec{x}_i}\ket{\vec{x}_j}\ket{f_K(\left|\vec{x}_i - \vec{x}_j\right|)},
\end{equation}
where $U_f$ is an efficient quantum circuit computing the $f_K$ function given the atomic positions $\vec{x}_i$ and $\vec{x}_j$, and $U_x$ is the circuit that loads the atomic positions to the quantum registers.
%It is worth mentioning that a wild idea can be implementing a trained classical neural networks (such as AlphaFold) using quantum gates, such that the structure of the protein can be predicted using the quantum circuit, and there's no need to manually load the atom positions. However, this is difficult in the foreseeable future due to the astronomical size of those neural networks.
$U_x$ can be implemented by QROM, which takes $\Or(\log N)$ circuit depth and $\Or(N\log N)$ gates:
\begin{equation}
    U_x \sum_i \alpha_i \ket{i}\ket{0,0,0} = \sum_i \alpha_i \ket{i}\ket{x_i,y_i,z_i}.
\end{equation}

%The matrix element $K_{i,j}$ between any pair of atoms $i,j$ can finally be computed by calling the atomic position oracle twice and the spring constant function oracle $U_f$, yielding the matrix element loading oracle:

%\subsection{Brute-force sparse access}
%If we wish to use the simulation algorithm proposed in Ref.~\cite{BBK23}, the access model of the $K$ matrix needs to be \emph{sparse}, which means that for any $i,k$, the matrix loading oracle should give $K_{i,j}$ where $j$ is the $k$th non-zero element of the $i$th row of $K$.
%The sparse access is the requirement to employ many efficient digital quantum simulation algorithms for sparse Hamiltonians.
%The data access in this case consists of $U_x$, $U_f$, and the \emph{sparse oracle} $U_s$ which is defined by

%Once $i$ and $j$ are known, one can apply $U_x$ and $U_f$ to compute $K_{i,j}$. 
Implementing $U_s$ is also a non-trivial task, although the method can be straightforward: one can compute $j_{i,k}$ classically for all $i,k$ pairs, and then implement the coherent access to all $(i,k)\mapsto j_{i,k}$ using QROM. The total amount of data to be stored is $\Or(N)$, which again requires $\Or(\log N)$ circuit depth and $\Or(N\log N)$ gates.
%$\Or(Nd)$ if $K$ is $d$-sparse.
We can thus claim that \prb{lowdepthmcl} is solved.
\begin{theorem}
\label{thm:lowdepthmcl}
    A sparse data access oracle to $K$ can be implemented in $\Or(\log N)$ depth using $\Or(N \log N)$ gates.
\end{theorem}
We notice that 
%due to the randomness in protein structure data, 
the naive algorithm for computing $j_{i,k}$ described above %involves calculating the distances between all pairs of atoms in the model, yielding a 
requires $\Or(N^2)$-time classical computation, while the information-theoretic lower bound is $\Omega(N)$. We also notice that a \emph{modification} of the sparse-access oracle is costly. We consider an example case where $K$ is $d$-sparse and the atom $i'$ is initially far away from all other atoms. If we want to change its position to make it within the reach of $d$ other atoms, then we must add $d$ entries to $j_{i',k}$ and modify up to $\Or(d^2)$ entries of $j_{i,k}$ for other indices $i$'s, since we may need to change the order of non-zero elements in its row. The same cost also applies to \emph{adding} or \emph{removing} atoms from the protein molecule. The $d^2$-dependence in the modification cost is acceptable when $K$ is sufficiently sparse. However, when the cut-off radius in GNM or ANM is large, $d$ can be significantly greater than $\polylog(N)$.

In fact, protein structure modification can be a useful feature: suppose we have built the QROM circuit for a protein and wish to load a slightly different one (with modified atom positions or atom numbers), then it is preferrable if the new molecule can be loaded by the same circuit with minimal updates, due to the $\Omega(N)$ cost of re-constructing the oracle for any protein. %This motivates us to consider the following problem.
\begin{problem}[Efficient modification of $K$]
\label{prb:efficientmodifymcl}
Build a $\Or(\polylog N)$-depth data access oracle for $K$ supporting efficient modification of the protein structure.
\end{problem}

%\subsection{A flexible binary tree data structure from Ref.~\cite{kerenidis2016quantum}}

%The above brute-force sparse oracle construction suffers from a huge cost of modification and lacks flexibility.

We are inspired by Ref.~\cite{kerenidis2016quantum} to use a different data structure based on a binary tree to allow for efficient modification of the protein structure as well as low-depth data loading. With this, we prove the following theorem to solve \prb{efficientmodifymcl}. The proof details can be found in S6 of Supplemental Materials 
\begin{theorem}
\label{thm:efficientmodifymcl}
There is a data access oracle of $K$ in $\Or(\polylog N)$ depth which supports protein structure modification (including atom adding, atom deletion, atom mass modification, and atom moving) by changing $\Or(d\cdot \polylog N)$ values stored in QROM.
\end{theorem}

%\jpl{Chunhao and Jin-Peng: you can add around 500 words for the simulation section}

%\vspace{0.2cm}
\subsubsection*{(iii) Simulating protein dynamics on quantum computers:}

In our problem formulation, the input and output
are quantum states \eqref{psi-encode} whose amplitudes encode the velocities and displacements of the protein molecules. We consider several widely used models for simulating protein dynamics and design quantum algorithms to perform these simulations efficiently.
%\\

%\jpl{Chunhao: please revise the simulation part and make it more like a letter}

%\jpl{Each problem should have its background and significance}

\paragraph{Harmonic oscillator model.} 
A direct model for molecular dynamics is based on Newton's equations of motion, corresponding to the microcanonical ensemble \cite{frenkel2023understanding}, where the dynamics occur in isolation. By applying the normal mode approximation \cite{bahar2010normal}, these equations can be written as a system of harmonic oscillators as follows:
\begin{equation}\label{eq:md-ham}
M\frac{\d^2\bm u(t)}{\d t^2} + K\bm u(t) = 0,
\end{equation}
with mass matrix $M$ and stiffness matrix $K$.
We consider a change of variables $\bm y(t) = \sqrt{M} \bm u(t)$. This allows us to rewrite the dynamics as $\ddot{\bm y}(t) + A\bm y(t) = 0$, where $A \coloneqq (\sqrt{M})^{-1}K(\sqrt{M})^{-1}$. 

To enable a direct quantum simulation of \cref{eq:md-ham}, we follow the incidence matrix approach~\cite{CJO19,BBK23}, and reshape the equations as the Schr\"odinger equation
\begin{equation}
\mathrm{i}\frac{\d}{\d t} \ket{\psi(t))} = H \ket{\psi(t))}, \qquad 
H = -
    \begin{bmatrix}
        0 & B \\
        B^\dag & 0 
    \end{bmatrix},
\end{equation}
%\chw{Shall we use the normal font for $i$?}\jpl{Good idea}
with $\ket{\psi(t))} \coloneqq \frac{1}{\sqrt{2E}} [\dot{\bm y}(t), \mathrm{i}B^{\dagger} \bm y(t)]^T$. Here we construct $B$ such that the first block of $H$ can satisfy $BB^{\dagger} = A$~\cite{CJO19,BBK23}, and $E$ is a normalizing factor related to the total energy.
Thus, the dynamics in \cref{eq:md-ham} can be readily simulated by  performing Hamiltonian simulation to propagate initial state $\ket{\psi(0))}$ to a target final state $\ket{\psi(T))}$ with $\Or \bigl(\sqrt{\kappa_A}T + \log(1/\epsilon) \bigr)$ query and gate complexity~\cite{LC17}. Here $\kappa_A$ is the condition number of $A$. More details refer to S7.1 of Supplement Materials.
%\\

\paragraph{Inhomogeneous harmonic oscillator model.}
We can also consider the normal mode dynamics under a mechanical force
\begin{equation}\label{eq:md-inhom}
    M\frac{\d^2\bm u(t)}{\d t^2} + K\bm u(t) = F,
\end{equation}
with an inhomogeneous force term $F$, which can be used to steer the molecular dynamics simulations \cite{isralewitz2001steered}. 
We cannot directly apply the standard Hamiltonian simulation for the non-unitary dynamics. Instead, we apply a quantum linear ODE solver to produce the final state $\ket{\psi(T))}$ or the history state $\ket{\Psi} = \frac{1}{\sqrt{N_t+1}} \sum_{k=0}^{N_t}|k\rangle|\psi(t_k)\rangle$ with $\Or \bigl( \sqrt{\kappa_A}T \mathrm{polylog}(1/\epsilon) \bigr)$ query and gate complexity~\cite{ALL23,ACL23}.  For the discussion of quantum ODE solvers for the final state or history state outputs, readers can refer to S7.2 and S7.3 of Supplement Materials.
%\\

\paragraph{Langevin dynamics.}
Another important setting for molecular simulations is the canonical ensemble  \cite{frenkel2023understanding}, where the system interacts with a heat reservoir. A common approach to incorporate stochastic effects is through Langevin dynamics, which in our case can be written as:
\begin{equation}\label{eq:md-lge}
    M\frac{\d^2\bm u}{\d t^2} +  \gamma\frac{\d\bm u}{\d t} + K\bm u + \sigma \xi(t) = 0.
\end{equation}
Here $\gamma>0$ is the friction coefficient from the interaction of the protein with the surrounding solvent. In addition, $\xi(t)$ is a white noise with independent entries acting on each velocity component. The noise amplitude is related to the damping coefficient according to the fluctuation-dissipation theorem $\sigma = \sqrt{2k_B T \gamma}$.

To formulate an efficient algorithm, we express \cref{eq:md-lge} as stochastic differential equations in the It\^o's form $\frac{\dd}{\dd t} \ket{\phi} = -(i H  + \gamma I)   \ket{\phi} + \Sigma \xi(t)$. We will simulate the covariance $\rho(t) =\mathbb{E}[\ket{\phi(t)}\bra{\phi(t)}]$. It\^o's Lemma implies the master equation
\begin{equation}\label{eq:master}
\frac{\dd}{\dd t} \rho = -i[H, \rho ] - \{\gamma I, \rho \} + \Sigma \Sigma^\dag
\quad \Longrightarrow \quad
\rho(t) = e^{tJ} \rho(0) e^{tJ^\dag} + \int_0^t e^{\tau J} \Sigma \Sigma^\dag e^{\tau J^\dag}  \d \tau.
\end{equation}
To prepare $\rho(t)$, we first sample from the probability distribution proportional to the traces of the two terms $e^{tJ} \rho(0) e^{tJ^\dag}$ and $\int_0^t e^{\tau J} \Sigma \Sigma^\dag e^{\tau J^\dag}  \d \tau$, and then prepare the first or the second term. Each term requires implementing the Kraus operator via scaled Hamiltonian evolution $e^{tJ} = e^{t(-iH-\gamma I)} = e^{-\gamma}e^{-iHt}$. Our algorithm, motivated by those for open quantum systems, produces $\rho(T)$ with $\Or \bigl( \sqrt{\kappa_A}T + \log(1/\epsilon) \bigr)$ query and gate complexity.  Readers may refer to S7.4 of Supplement Materials for more details.

\begin{table}[htpb]
\renewcommand{\arraystretch}{1.5}
    \centering
    \scriptsize{
    \begin{tabular}{|c|c|c|c|c|c|}
      \hline\hline
      \textbf{Problem} & \textbf{Algorithm} &  \textbf{Complexity} \\
      \hline
      Harmonic oscillator & Hamiltonian simulation   & $\Or \Bigl(\sqrt{\kappa_A}T + \log(1/\epsilon) \Bigr)$ \\
      \hline
      Inhomogeneous harmonic oscillator & Linear combination of Hamiltonian simulation &   $\Or \Bigl(\sqrt{\kappa_A}T \log^{1+1/\beta}(1/\epsilon) \Bigr)$ \\
      \hline
      History-state model & Quantum linear ODE solver   & $\Or \Bigl( \sqrt{\kappa_A}T \polylog(1/\epsilon) \Bigr)$ \\
      \hline
      Langevin dynamics  & Open quantum system simulation &   $\Or \Bigl(\sqrt{\kappa_A}T + \log(1/\epsilon) \Bigr)$ \\
      \hline\hline
    \end{tabular}
    }
    \caption{Summary of quantum simulation algorithms for the protein dynamics. Here $\kappa$ is the norm of $A = (\sqrt{M})^{-1}K(\sqrt{M})^{-1}$, $T$ is the evolution time, and $\beta\in(0,1)$. 
    %\ZL{where is $n$ in the table?}\jpl{polylog scaling of the dimension can be ignored in this table?}\ZL{I guess so, but we explicitly said "here $n$ is the number of protein molecules" but didn't use $n$ at all. I mean we should delete this sentence.}\jpl{I see}
    }
    \label{tab:simulation}
\end{table}

Higher order moments can be simulated as well.  \cref{tab:simulation} summarized the quantum algorithms with the corresponding complexity. 

%\jpl{Jin-Peng: you can add around 500 words for the simulation section}

%\vspace{-0.6cm}
\subsubsection*{(iv) Computing properties of proteins:} 

As in classical simulations, the goal of our quantum algorithms is to compute certain properties of proteins within a given error tolerance $\epsilon$. %In our paper, we consider two types of input models:
We assess the complexity associated with the estimation of these properties in terms of  two types of input models~\cite{DLT22}:
\\ 
(a) block-encoding (BE) model: let $U$ be an $\alpha$-block-encoding of the Hamiltonian $H$;\\
(b) Quantum evolution (QE) model: there is a quantum algorithm $U(t)$ that prepares an amplitude-encoded solution $|\psi(t)\rangle$ at time $t$. The QE model extends the Hamiltonian evolution (HE) model when $U(t)$ is generated by a Hamiltonian~\cite{DLT22}. 
%\xl{add references to these two input models. }

We discuss various types of application models with queries to the input model of $H$, and a Hermitian matrix $O$ that represent physical properties as an quantum expectation, such as kinetic and potential energy, low vibration modes, density of states and Chebyshev moments, correlation and root mean square of displacement, and optimal control of molecular dynamics. We summarize the applications in in Table~\ref{tab:application}. The detailed problem models and quantum algorithms of computing protein properties refer to S8 of Supplement Materials.

\begin{table}[H]
\renewcommand{\arraystretch}{1.5}
    \centering
    \scriptsize{
    \begin{tabular}{|c|c|c|c|c|c|c|}
      \hline\hline
      \textbf{Problem} & \textbf{Algorithm} & \textbf{Complexity} & \textbf{Model} \\
      \hline
      Kinetic and potential energy & HS and AA & $\Or(1/\epsilon)$ & QE \\
      \hline
      Low vibration modes & HS and AA & $\Or\bigl( 1/(\gamma\epsilon) \bigr)$ & QE \\
      \hline
      Density of States, Chebyshev moments & QSVT and AA & $\Or(\alpha/\epsilon^2)$, $\Or(\alpha k/\epsilon)$ & BE \\
      \hline
      Correlation, root mean square displacement & QSVT and AA & $\Or(\alpha /\epsilon)$ & BE \\
      \hline
      Molecular dynamical control & HS and AA & $\Or(\beta/\epsilon)$ & QE \\
      \hline\hline
    \end{tabular}
    }
    \caption{Summary of end-to-end quantum applications for the protein dynamics. Here $\epsilon$ is the error tolerance, $\alpha, \beta$ are the block-encoding factors of Hamiltonian $H$ and observable $O$, $\gamma$ is the overlap between initial and ground states, $k$ is the degree of the Chebyshev moment. HS: Hamiltonian simulation. AA: amplitude amplification. QSVT: quantum singular value transformation. QE: quantum evolution model. BE: block-encoding model.}
    \label{tab:application}
\end{table}

\subsection*{II.Numerical Experiments}

\begin{figure}[htp]
    \centering
    \includegraphics[scale=0.2]{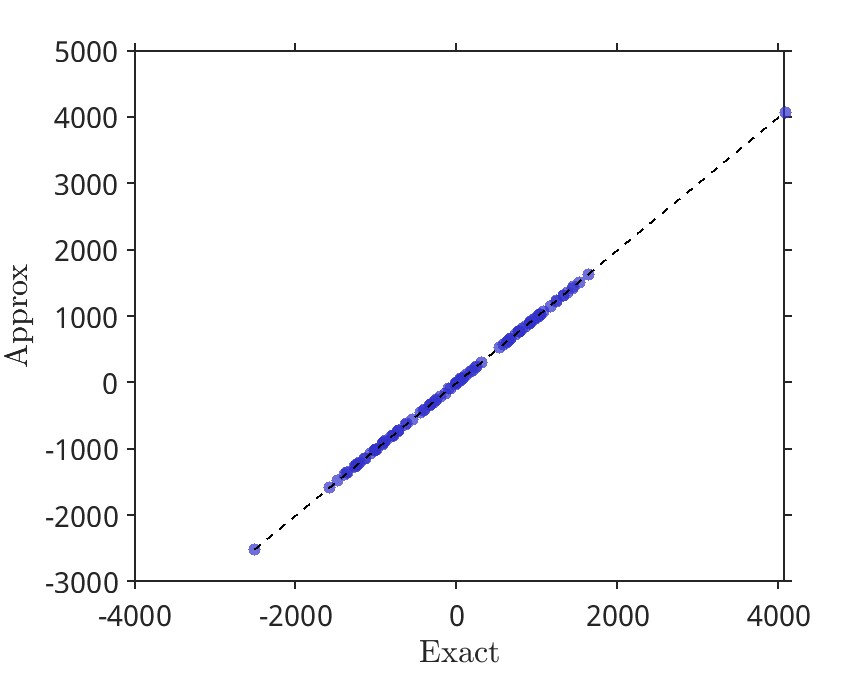}
    \includegraphics[scale=0.2]{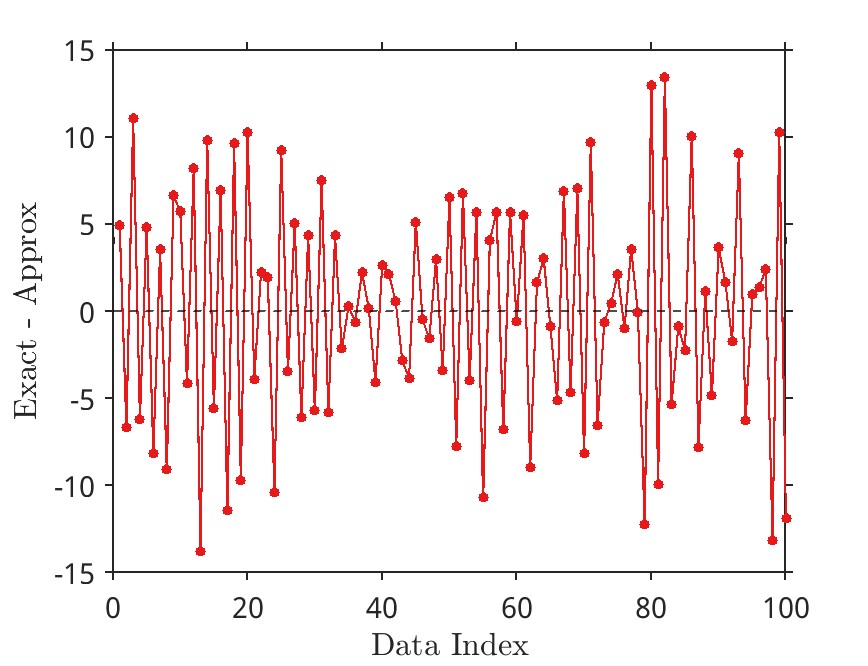}
    \caption{ (Left) Comparison of 100 moments from exact and approximate calculations. Each marker represents a pair \((\text{Exact},\ \text{Approx})\), with the dashed line indicating perfect agreement \((y = x)\); (Right) Deviations of approximate from exact moments (on the orders of $10^2$ and $10^3$), shown by plotting \(\text{Exact} - \text{Approx}\) against the moment index. The horizontal dashed line at zero provides a visual reference for the absence of error. }
    \label{fgr:cmp}
\end{figure}

\begin{figure}[htp]
    \centering
    \includegraphics[scale=0.25]{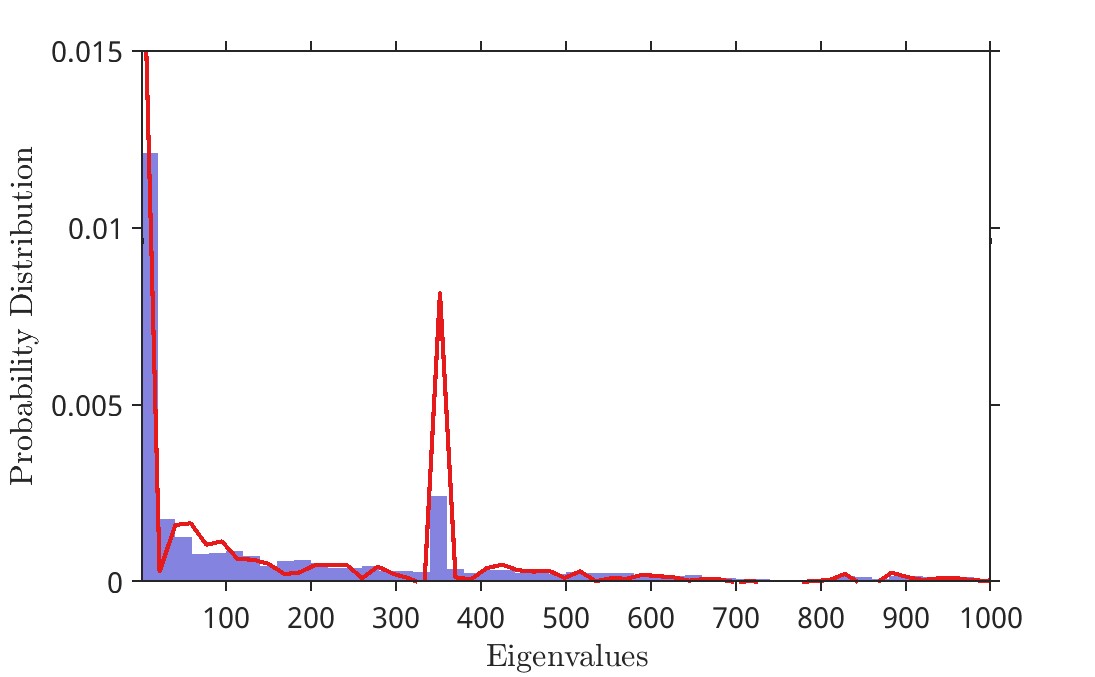}
    \caption{ Comparison of estimated the density of states with the histogram of the eigenvalues computed directly from $H$.}
    \label{fgr:cmp-hist}
\end{figure}

We designed a series of numerical experiments to evaluate our algorithms, conducting these tests on classical devices to demonstrate their feasibility.  In the first experiment, we tested our algorithm for estimating the density of states, 
\begin{equation}\label{eq: dos}
    \varrho(\lambda) = \frac{1}{N}\sum_{j=0}^{N-1} \delta(\lambda-\lambda_j),
\end{equation}
which provides direct insights into the distribution of vibrational modes~\cite{kamp2005spectral}.
As a case study, we considered the Crambin protein, consisting of 1360 atoms. Since our method relies on the generalized moment approach, we began by computing the exact moment values through brute-force matrix multiplications. These exact values were then compared with the moments estimated using traces of the Chebyshev polynomials $\mu^{\varrho}_k \approx \Tr\Bigl(\frac{I}{N}T_k(H/\alpha)\Bigr)$, as shown in \cref{fgr:cmp}.
Using the estimated moments, we reconstructed the density of states and directly compared it to the histogram of the eigenvalues. As shown in \cref{fgr:cmp-hist}, the reconstructed density aligns remarkably well with the eigenvalue histogram, demonstrating excellent agreement.

%\begin{figure}[htp]
%    \centering
%    \begin{subfigure}[t]{0.5\textwidth}
%        \centering
%        \includegraphics[scale=0.1]{numerical-test/moments.jpg}
%        \caption{Comparison of the exact moments and the approximate moments computed from random samples in \eqref{mu_n}}
%    \end{subfigure}%
%    \begin{subfigure}[t]{0.5\textwidth}
%        \centering
%        \includegraphics[scale=0.2]{numerical-test/hist-cram.jpg}
%        \caption{Comparison of the density of states}
%    \end{subfigure}
%    \caption{}
%\end{figure}
\iffalse 

In our second test, we follow the approach in \cite{arkun2012combining} and study the folding mechanism via an LQR control with an infinity horizon. We simulate the dynamics of \eqref{eq: lqr1}. Figure \ref{fig:lqr-energy} shows the time history of the total energy. 

%\vspace{-2.5em}
\begin{figure}[htp]
    \centering
    \includegraphics[scale=0.11]{ch1.jpg} $\to$
    \includegraphics[scale=0.11]{ch2.jpg} $\to$
    \includegraphics[scale=0.11]{ch3.jpg} $\to$
    \includegraphics[scale=0.11]{ch4.jpg}
    \includegraphics[scale=0.15]{lqr-energy.jpg}
    \caption{(Left) the structural changes under the control;
    (Right) The total energy $\bm y(t)^T  Q \bm y(t)$ following the optimal control 
    $\bm u(t)$. }
    \label{fig:lqr-energy}
\end{figure}
\fi 

In our second experiment, we consider the Linear-Quadratic Regulator (LQR) control of molecular dynamics. The goal is to determine the control signals $\bm u(t)$ to minimize the cost function,
\begin{equation}\label{eq: lqr1}
   M\frac{\d^2 \bm u}{\d t^2} + \gamma \frac{\d \bm u}{\d t} + K\bm u = \bm f(t), \qquad J[\bm f]=  \frac{1}{2}\bm u(T)^T S \bm u(T) + \int_0^{+\infty} \bm u(t)^T  Q \bm u(t) + \bm f(t)^T  R \bm f(t) dt.
\end{equation}
The solution provides the optimal feedback $\bm f(t)= -K \bm u(t)$ with optimal $K$.

\begin{figure}[htp]
    \centering
     \includegraphics[scale=0.18]{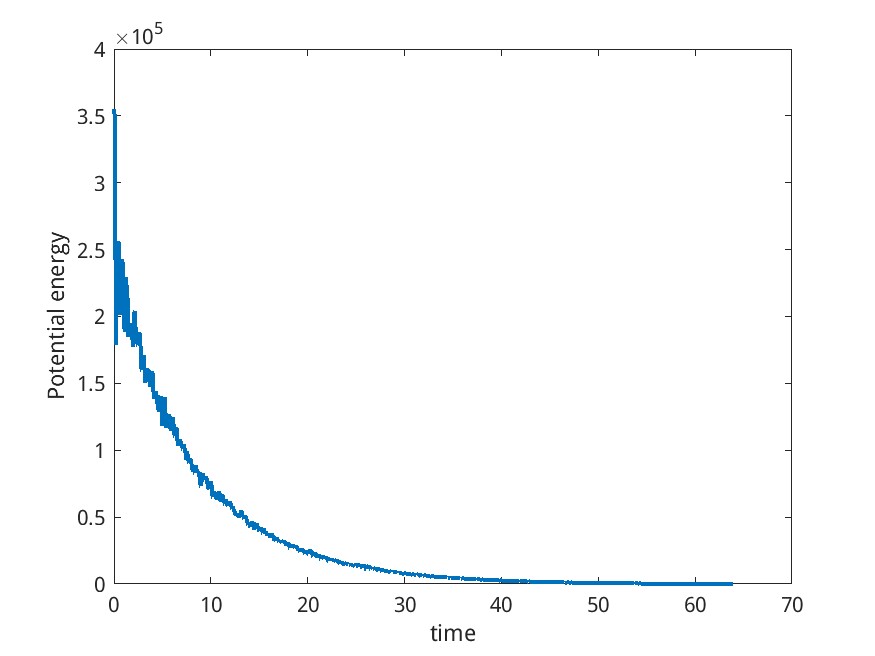}
    \includegraphics[scale=0.13]{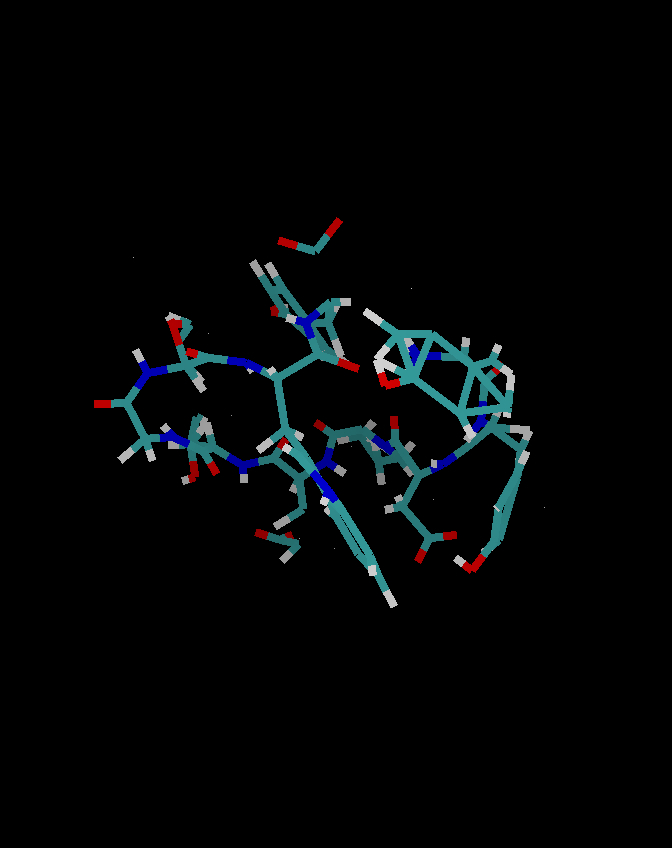} $\to$
    \includegraphics[scale=0.13]{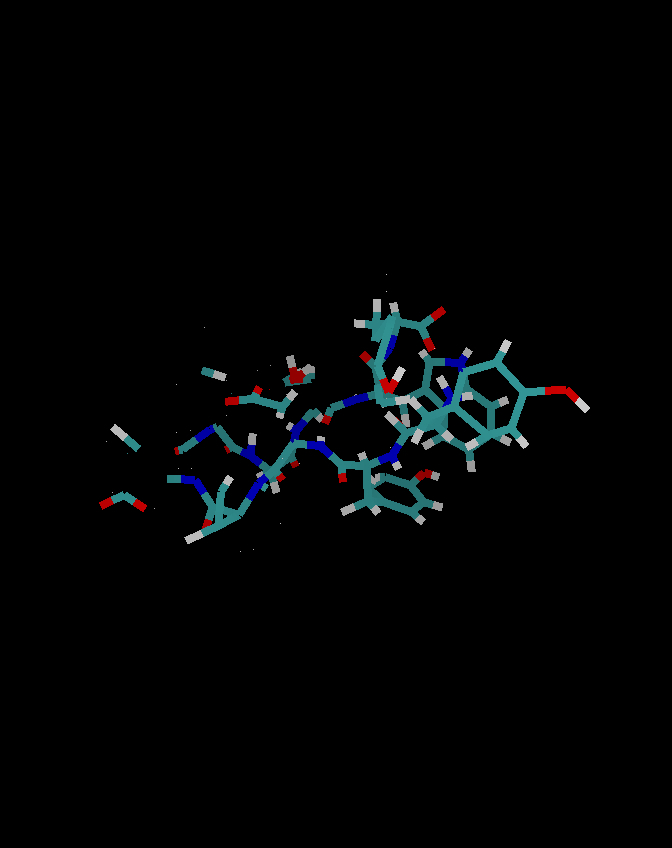} $\to$
    \includegraphics[scale=0.13]{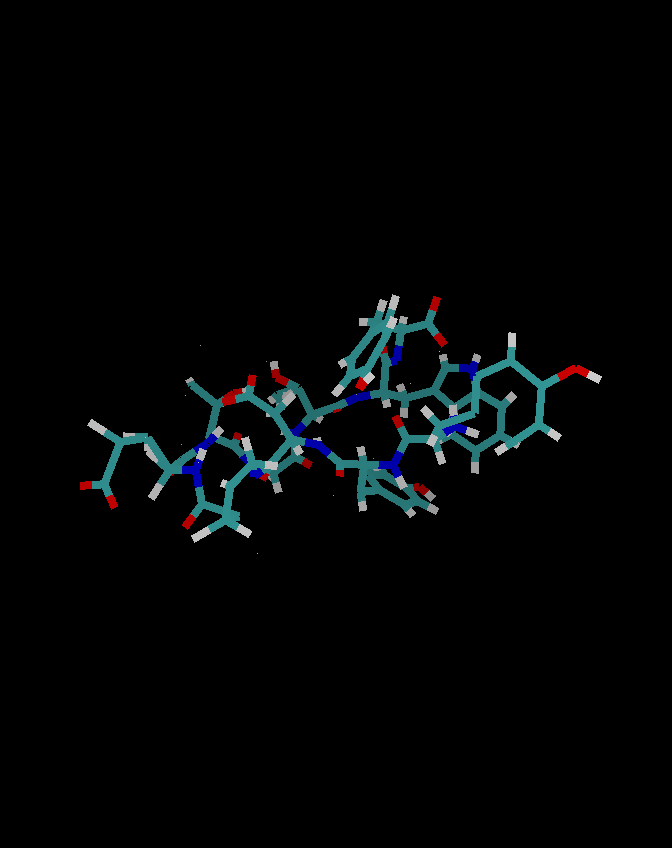} 
    \caption{(L) The total energy following the optimal control 
    $\bm u(t)$; (R) the structural changes under the control;  }
    \label{fig:lqr-energy1}
\end{figure}

We follow the approach in \cite{arkun2012combining} and study the folding mechanism via a LQR control.
To demonstrate the effect of the control, we consider a Chignolin protein and starting configuration away from equilibrium, followed by a control algorithm that minimizes the energy \cite{arkun2012combining}.  
\cref{fig:lqr-energy1} shows the resulting potential energy in time, has has reached the minimum values. The corresponding displacement from the controlled dynamics in \cref{eq: lqr1} has been mapped back to the protein configurations, as shown the snapshots in  \cref{fig:lqr-energy1}, indicating that the molecule has been guided back to equilibrium. 

%\vspace{-1.5em}

\subsection*{Discussion and outlook}

In this work, motivated by the critical role of protein dynamics simulations in biology, we conducted a systematic study of quantum computing algorithms for simulating protein dynamics and the resulting properties. Our focus is end-to-end quantum algorithms for simulating various protein dynamical models. We present a detailed analysis of the computational complexity of these quantum simulation algorithms, as well as the quantum protocols for data input (read-in) and output (read-out). Additionally, we discuss several concrete applications and illustrate the potential efficiency of these algorithms through classical numerical experiments.

\section*{Data availability statement}
The full data for this work is available at...

\section*{Code availability statement}

The full code for this work is available at...

\section*{Acknowledgments}
We thank Dong An, Andrew Childs, and Lin Lin for valuable discussions. ZL acknowledges support from the National Science Foundation (QLCI grant OMA-2120757). JPL acknowledges support from Tsinghua University and Beijing Institute of Mathematical Sciences and Applications. 

\section*{Author contributions statement}

All authors have contributed to pursuing the theoretical analysis and to writing this manuscript.

\section*{Competing interests statement}

There are no competing interests.

\bibliographystyle{abbrv}
\bibliography{ref}

\newpage

\includepdf[pages=-,pagecommand={},trim=0 2cm 0 2cm,clip]{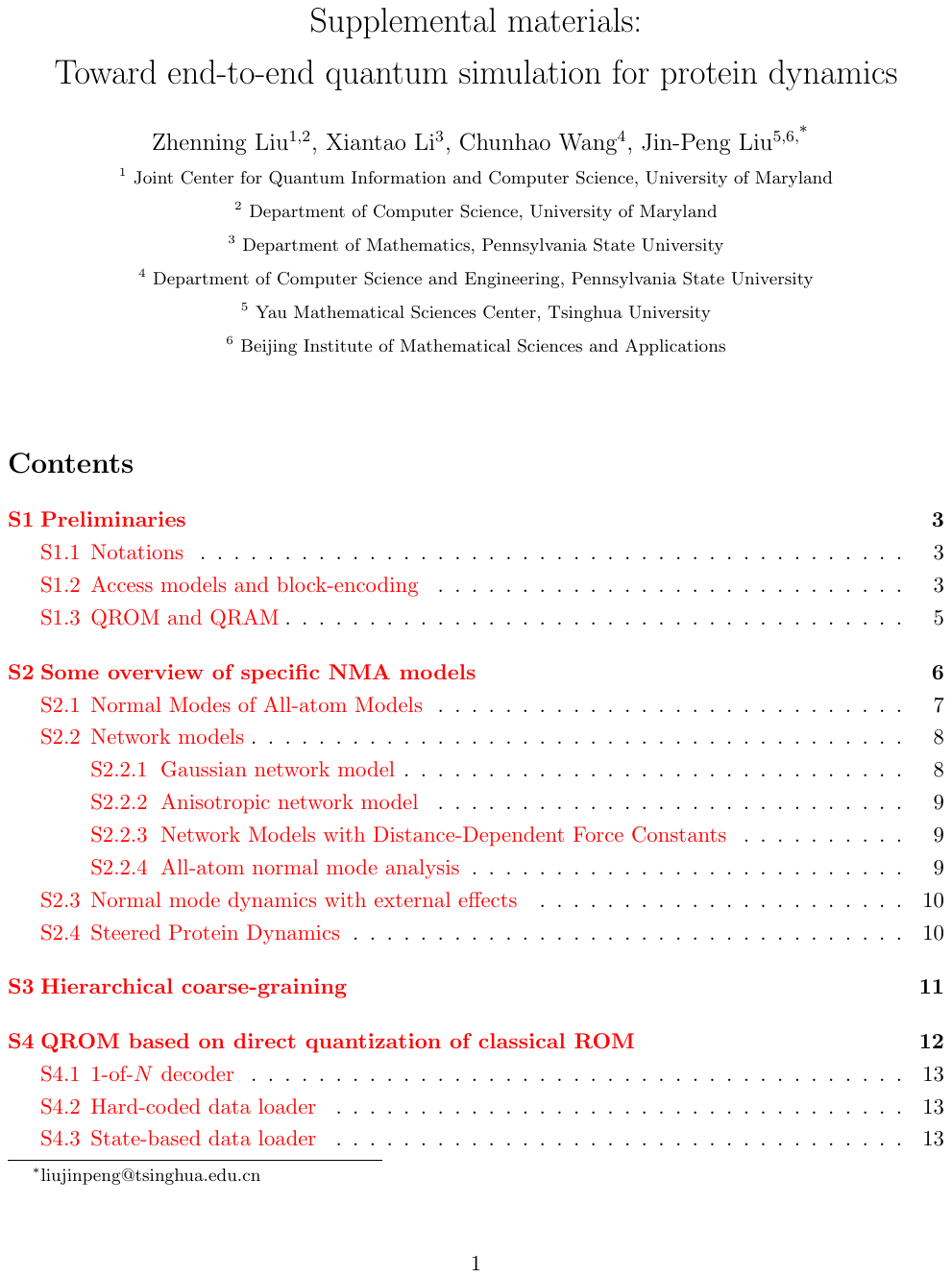}

\end{document}